\documentclass[prl,twocolumn,superscriptaddress,floatfix]{revtex4-2}

\usepackage{amsmath,amssymb}
\usepackage{graphicx}
\graphicspath{{fig/}{../eftdft-note/fig/}}
\usepackage{dcolumn}
\usepackage{bm}
\usepackage{booktabs}
\usepackage{xcolor}
\usepackage{hyperref}

\newcommand{\bk}{\mathbf{k}}

\newcommand{\bq}{\mathbf{q}}

\newcommand{\br}{\mathbf{r}}

\newcommand{\bK}{\mathbf{K}}
\newcommand{\bG}{\mathbf{G}}

\newcommand{\barpsi}{\bar{\psi}}

\newcommand{\calD}{\mathcal{D}}
\newcommand{\bra}[1]{\langle #1 |}
\newcommand{\ket}[1]{| #1 \rangle}

\newcommand{\eint}[2]{\langle #1 || #2 \rangle}
\newcommand{\Ha}{\text{Ha}}
\newcommand{\eV}{\text{eV}}
\newcommand{\Bohr}{\text{Bohr}}

\newcommand{\prlauthors}{%
  \author{Xiansheng Cai}
  \affiliation{Institute of Theoretical Physics, Chinese Academy of Sciences, Beijing 100190, China}
  \author{Han Wang}
  \affiliation{Institute of Theoretical Physics, Chinese Academy of Sciences, Beijing 100190, China}
  \author{Kun Chen}
  \email{chenkun@itp.ac.cn}
  \affiliation{Institute of Theoretical Physics, Chinese Academy of Sciences, Beijing 100190, China}
}

\usepackage{feynmp-auto}  % needed for inline Feynman diagrams in combined file

\begin{document}

\title{Kohn--Sham Hamiltonian from Effective Field Theory:\\
Quasiparticle Band Narrowing from Frozen Core Dynamics}

\prlauthors
\date{\today}

% ═══════════════════════════════════════════
% Main text
% ═══════════════════════════════════════════
\begin{abstract}
    Kohn--Sham (KS) eigenvalues are routinely compared with
    angle-resolved photoemission (ARPES) and used as input for many-body
    methods, yet density functional theory (DFT) assigns them no
    physical meaning.  For alkali and alkaline-earth metals, KS
    bandwidths overestimate ARPES measurements by 20--35\%, a
    discrepancy that persists across all exchange-correlation
    functionals.  We construct an effective field theory (EFT) of the
    inhomogeneous electron gas and show that two conditions imply
    KS bands \emph{are} the quasiparticle bands, up to a frozen-core
    renormalization factor $z^{\mathrm{core}}$: a scale separation
    between core excitation energies and the valence Fermi energy, and
    an approximate Galilean invariance of the uniform electron gas
    confirmed by diagrammatic Monte Carlo.  This factor reflects
    dynamical core excitations that conventional pseudopotentials freeze
    out and no static potential can capture.  The correction
    $1 - z^{\mathrm{core}}$ reaches 20--35\% for alkali metals but
    falls below 5\% for Al and Si, explaining both the failure and
    success of KS band theory.  We derive a closed-form post-SCF
    formula and validate it for Li, Na, K, Ca, Mg, Al, and Si;
    the predicted quasiparticle bands resolve the long-standing
    ARPES bandwidth discrepancy, matching embedded dynamical
    mean-field theory at negligible cost.
    This work also exemplifies first-principles agentic science,
    a direction particularly suited to the AGI-for-Science
    paradigm~\cite{cai2025lac}: an LLM-co-developed derivation
    with controlled approximations, verified symbolically and
    against a few experiments, becomes a deterministic harness
    for agentic scale-out, resolving simultaneously the LLM audit
    bottleneck and the non-falsifiability of fit-based
    AI-for-science.
\end{abstract}

\maketitle

\begin{figure*}[t]
    \centering
    \includegraphics[width=0.95\textwidth]{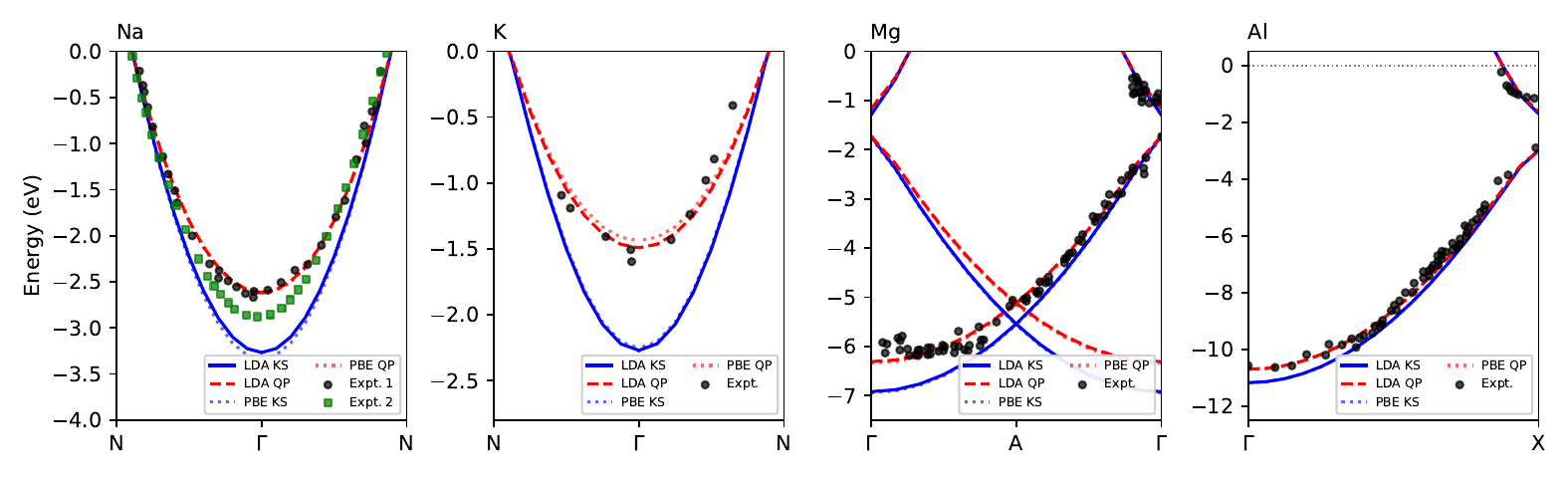}
    \caption{\label{fig:arpes_bands}
        Valence bands compared with ARPES data. Solid: LDA KS (blue) and
        QP (red). Dotted: PBE KS and QP.
        Na and K along N--$\Gamma$--N:
        Expt.~1 = Ref.~\cite{lyo1988}, Expt.~2 = Ref.~\cite{potorochin2022},
        K expt.\ = Ref.~\cite{itchkawitz1990}.
        Mg along $\Gamma$--A--$\Gamma$: expt.\ = Ref.~\cite{bartynski1986}.
        Al along $\Gamma$--X: expt.\ = Ref.~\cite{levinson1983}.
        PBE closely tracks LDA, confirming robustness to the functional.}
\end{figure*}

The Kohn--Sham (KS) Hamiltonian is the workhorse of modern
electronic structure theory.  For $sp$-bonded metals and
semiconductors, its eigenvalues successfully approximate
quasiparticle energies measured by angle-resolved photoemission
spectroscopy (ARPES)~\cite{levinson1983,grobman1972}, and this
agreement is widely taken for granted.  Yet within density
functional theory (DFT) the KS eigenvalues are purely
auxiliary~\cite{perdew1982,sham1983}: there is no formal guarantee
that they should approximate quasiparticle energies at all.
For alkali and alkaline-earth metals this gap between practice and
theory becomes visible: KS bandwidths systematically overestimate
ARPES measurements by
20--35\%~\cite{lyo1988,itchkawitz1990,potorochin2022}, a
discrepancy first noted in the 1980s that persists regardless of
the exchange-correlation functional~\cite{mandal2022}.  That only
all-electron embedded dynamical mean-field theory
(eDMFT)~\cite{mandal2022}, a many-body method that treats
dynamical correlations beyond DFT, resolves this discrepancy
signals that the missing physics is not a better functional but
something outside the KS framework altogether.

What is missing is a first-principles theory that establishes when
and why KS eigenvalues approximate quasiparticle energies.
DFT itself is not at fault: it provides a rigorous variational
principle for the ground-state
density~\cite{hohenberg1964,kohn1965}.  The gap lies in the step from DFT to the KS Hamiltonian,
whose eigenvalues lack physical justification yet serve as
the starting point for ARPES interpretation, $GW$, DMFT, and
DFT+$U$.

Starting from the Born--Oppenheimer action for electrons in a
lattice potential with Coulomb interactions, we construct an
effective field theory (EFT) of the inhomogeneous electron gas,
guided by two inputs: (i) a scale separation between core
excitation energies and the valence Fermi energy that allows core
electrons to be integrated out, and (ii) locality properties of the
interacting uniform electron gas (UEG) at metallic densities,
established by high-order diagrammatic Monte Carlo (DiagMC) calculations,
that dictate how many-body effects are absorbed into renormalized
parameters.  The tree-level quasiparticle propagator of this EFT
turns out to be
\begin{equation}\label{eq:G_factor}
    G^{-1}_{\nu\bk}(i\omega) \approx
    (z^{\mathrm{val}})^{-1}\bigl[
        (z^{\mathrm{core}}_{\nu\bk})^{-1}\,i\omega
        - (\epsilon^{\mathrm{KS}}_{\nu\bk} - \epsilon_F)\bigr],
\end{equation}
where, remarkably, the quasiparticle dispersion is governed by
the eigenvalues $\epsilon^{\mathrm{KS}}_{\nu\bk}$ of the
Kohn--Sham Hamiltonian
$\hat H_{\mathrm{KS}} = -\nabla^2/2 + V_{\mathrm{PSP}} + V_H[n]
    + V_{xc}[n]$.  Here $z^{\mathrm{val}}$ is the valence
quasiparticle residue and $z^{\mathrm{core}}_{\nu\bk}$ is a
closed-form frozen-core renormalization determined by the core
form factors $f^c_K$ and excitation energies $\Delta E_c$, given
explicitly by Eq.~\eqref{eq:zcore_nk} below.  In
Eq.~\eqref{eq:G_factor}, $z^{\mathrm{val}}$ is an overall prefactor
that does not shift the pole, while $z^{\mathrm{core}}_{\nu\bk}$
rescales the frequency relative to the dispersion---an effective
time dilation caused by virtual core excitations.  Locating the
pole gives the quasiparticle energy measured by ARPES,
\begin{equation}\label{eq:qp_energy}
    \epsilon^{\mathrm{QP}}_{\nu\bk} \approx \epsilon_F
    + z^{\mathrm{core}}_{\nu\bk}\,
    \bigl(\epsilon^{\mathrm{KS}}_{\nu\bk} - \epsilon_F\bigr),
\end{equation}
so that the Kohn--Sham eigenvalues \emph{are} the quasiparticle
energies, compressed toward $\epsilon_F$ by the state-dependent
factor $z^{\mathrm{core}}_{\nu\bk}$.
Equation~\eqref{eq:G_factor} is not an ansatz: it is the unique
tree-level consequence of conditions (i) and (ii).  The frozen-core
factor $z^{\mathrm{core}}$, absent
from DFT and overlooked by conventional pseudopotential
theory~\cite{phillips1959,pickett1989}, produces a correction
$1 - z^{\mathrm{core}}$ of $20$--$35\%$ in alkali and
alkaline-earth metals (where $\Delta E_c$ is only a few Hartree)
but below $5\%$ in $sp$-bonded metals and semiconductors,
directly explaining the bandwidth discrepancy introduced above.
We now describe how each condition determines a different
aspect of the result: (i)~produces the dynamical correction
$z^{\mathrm{core}}$, while (ii)~determines the static
Hamiltonian $\hat H_{\mathrm{KS}}$.

Condition~(i) is the scale separation between the core
excitation energies $\Delta E_c \sim 2$--$100\;\Ha$ and the
valence Fermi energy $\epsilon_F \sim 0.1\;\Ha$---three orders of
magnitude apart.  It is handled by a dual-fermion
transformation~\cite{rubtsov2008,rubtsov2012,ribic2017}, which integrates
out the core in a controlled expansion with small parameter
$\epsilon_F/\Delta E_c \lesssim 0.05$.  The result is a
valence-only propagator whose static part reproduces a
conventional
pseudopotential~\cite{hamann1979,troullier1991,goedecker1996} and
whose dynamical part is separable in Bloch momenta.  Solving the
Dyson equation for this separable self-energy gives the
closed-form frozen-core renormalization appearing in
Eqs.~\eqref{eq:G_factor}--\eqref{eq:qp_energy},
\begin{equation}\label{eq:zcore_nk}
    z^{\mathrm{core}}_{\nu\bk} = \frac{1}{1 + \sum_c
    \bigl|F^c_{\nu\bk}\bigr|^2 / \Delta E_c^2}, \quad
    F^c_{\nu\bk} = \sum_{\bG} c_{\nu\bk}(\bG)\,
    f^c_{|\bk + \bG|}\,,
\end{equation}
where $f^c_K$ is the core form factor
    [Eq.~\eqref{eq:form_factor}], $\Delta E_c$ the core excitation
energy of channel $c$, and $c_{\nu\bk}(\bG)$ the Bloch coefficients
of the valence eigenstates.

Condition~(i) determines the dynamical correction but not
the static theory.  That the tree-level Hamiltonian takes
the KS form requires condition~(ii), the approximate
Galilean invariance.  DiagMC data for the UEG at metallic
densities (Fig.~\ref{fig:ueg_sigma}) show that the
interacting self-energy depends on $(i\omega,\bk)$ approximately
only through the Galilean combination $i\omega - k^2/(2m)$: the
quasiparticle residue $z(k)$ is nearly constant and $m^*\approx m$
throughout the occupied Fermi ball.  This means that the insights
of the Polchinski--Shankar Fermi-surface
RG~\cite{polchinski1992,shankar1994}---in particular the hierarchy
of relevant, marginal, and irrelevant couplings---extend
approximately to the entire Fermi sea, directly prescribing how
many-body effects should be absorbed into renormalized parameters.
The resulting renormalized perturbation theory (RPT) for the
inhomogeneous valence liquid yields the tree-level quasiparticle
propagator Eq.~\eqref{eq:G_factor} and thereby explains the
origin and physical meaning of the KS Hamiltonian.  The
$(z^{\mathrm{core}})^{-1}$ factor survives because core electrons
do not screen dynamically ($\Gamma_3^{\mathrm{core}} \approx 1$,
polarizability suppressed by
$\epsilon_F/\Delta E_c \sim 0.02$--$0.05$).
We validate Eq.~\eqref{eq:qp_energy} against eDMFT and ARPES for
Li, Na, K, Ca, Mg, Al, and Si
(Table~\ref{tab:comparison}, Fig.~\ref{fig:arpes_bands}).
Both this first-principles reconstruction and its seven-element
validation were produced in close collaboration with LLM-based
coding agents, exemplifying a workflow we call first-principles
agentic science (see End Matter).

\textit{Integrating out the core.---}The first EFT step starts
from the electron--ion Lagrangian
($S = \int_0^\beta d\tau\, L$),
\begin{equation}\label{eq:L_BO}
    L = \barpsi\, g_0^{-1}\, \psi + L_{\mathrm{int}},
\end{equation}
where $g_0^{-1} = \partial_\tau - \nabla^2/2 + V_{\mathrm{lat}}
    - \mu$ is the bare single-particle propagator,
$L_{\mathrm{int}} = \frac{1}{2}\int_{\br\br'}
    \barpsi_\br \barpsi_{\br'}\, v(\br{-}\br')\,
    \psi_{\br'} \psi_\br$ is the Coulomb interaction, and
$\barpsi A \psi \equiv \int_{\br\br'} \barpsi_\br\,
    A(\br,\br')\, \psi_{\br'}$.
The hierarchy $\epsilon_F \ll \Delta E_c$ gives a controlled
expansion with small parameter
$\epsilon_F/\Delta E_c \lesssim 0.05$: integrating out the core
Grassmann fields via a dual-fermion
transformation~\cite{rubtsov2008,rubtsov2012,ribic2017}
(see Supplemental Material~\cite{supp})
produces a valence effective action $S_{\mathrm{val}}$,
\begin{equation}\label{eq:L_val}
    L_{\mathrm{val}} = \barpsi\,
    (g^{\mathrm{val}})^{-1}\, \psi
    + L_{\mathrm{int}}^{\mathrm{val}},
\end{equation}
with bare valence propagator
$(g^{\mathrm{val}})^{-1}
    = (z^{\mathrm{core}})^{-1}\partial_\tau - \nabla^2/2
    + V_{\mathrm{PSP}} - \mu$
(where $V_{\mathrm{PSP}}$ is the nonlocal static
pseudopotential~\cite{hamann1979,troullier1991,goedecker1996,vanderbilt1990}),
$L_{\mathrm{int}}^{\mathrm{val}}$ the Coulomb interaction among
valence electrons, and a dynamical part.  The full
pseudopotential splits as
\begin{equation}\label{eq:dvpp}
    (\delta V_{pp})_{st} = \Sigma^{\text{HF}}_{st}
    + \sum_c \frac{M_{sc}\, M_{tc}^*}{i\omega + \Delta E_c}
    + O(\epsilon_F^2/\Delta E_c),
\end{equation}
where $s,t$ label valence orbitals, $\Sigma^{\text{HF}}$ is the
static Hartree--Fock self-energy of the core (reproducing
conventional norm-conserving
PSPs~\cite{hamann1979,troullier1991,goedecker1996,vanderbilt1990}),
and the second term has poles at the core excitation
energies $\Delta E_c$ with matrix elements $M_{sc}$ coupling
valence orbital $s$ to core hole $c$
(see Supplemental Material~\cite{supp}).
Its value at $\omega = \epsilon_F$ is a static shift already
absorbed into fitted PSP parameters; only the frequency
\emph{dependence} around $\epsilon_F$ is new, and this is
what produces $z^{\mathrm{core}}$.

In momentum space the frequency-dependent part is separable,
$V_{\text{dyn}}(\bK,\bK';i\omega) = \sum_c f^c_K\, f^c_{K'}/
    (i\omega + \Delta E_c)$, with the Wilson coefficient
\begin{equation}\label{eq:form_factor}
    f^c_K = \frac{\sqrt{4\pi}}{K}\int_0^\infty \!\!
    u_c(r)\,\bigl[V_{H,c}(r) - J_c\bigr]\,\sin(Kr)\,dr
\end{equation}
computed from atomic Kohn--Sham core wavefunctions
(see Supplemental Material~\cite{supp}).  Here $u_c(r) = r R_c(r)$ is the
reduced radial core wavefunction, $V_{H,c}$ its orbital Hartree
potential, and $J_c$ the self-Coulomb integral.  Projecting onto
Bloch eigenstates gives the coherent form factor
$F^c_{\nu\bk}$ of Eq.~\eqref{eq:zcore_nk}; solving the Dyson
equation for the resulting separable self-energy yields
$z^{\mathrm{core}}_{\nu\bk}$
(see Supplemental Material~\cite{supp}).
Figure~\ref{fig:delta_K} shows the dimensionless ratio
$f^c_K/\Delta E_c$ that controls the magnitude of the frozen-core
correction via
$1 - z^{\mathrm{core}}_\Gamma \approx (f^c_0/\Delta E_c)^2$ near
$\Gamma$.

\begin{figure}[t]
    \centering
    \includegraphics[width=0.95\columnwidth]{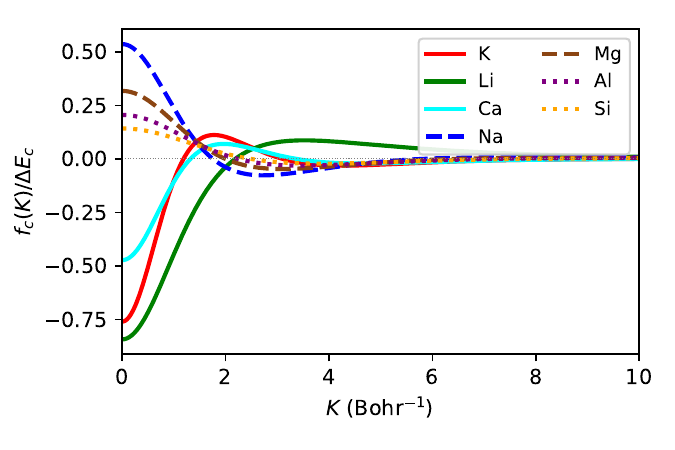}
    \caption{\label{fig:delta_K}
    Dimensionless ratio $f^c_K/\Delta E_c$ for the dominant core channel
    of each element.  This ratio controls $z^{\mathrm{core}}$
    [Eq.~\eqref{eq:zcore_nk}]: elements with large $|f^c/\Delta E_c|$
    have large corrections.  K and Li have the largest ratios because
    $\Delta E_c$ is small ($1.7$ to $2.6\;\Ha$) while $|f^c_K| \sim J_c
        \sim 1$ to $2\;\Ha$ is comparable.  The [Ne]-core series
    Na--Mg--Al--Si shows monotonically decreasing amplitude: $|f^c_K|$
    is nearly the same for all four because $J_c$ and the orbital size
    $\langle r\rangle_c$ scale inversely with $Z$, keeping
    $|f_c| \sim J_c \langle r\rangle_c \sim 1\;\Ha$; meanwhile
    $\Delta E_{2s}$ roughly triples from Na to Si.
    The sign change near $K \approx 2$ to $3\;\Bohr^{-1}$ leads to
    partial cancellation in the coherent sum $F^c_{\nu\bk}$.}
\end{figure}

\textit{From the valence EFT to Kohn--Sham.---}The second step
constructs a renormalized perturbation theory (RPT) for the
valence EFT, Eq.~\eqref{eq:L_val}.  The Polchinski--Shankar
RG~\cite{polchinski1992,shankar1994} for a Fermi liquid classifies
couplings into three categories: the local chemical potential is
\emph{relevant} (must be matched exactly), the quasiparticle
residue $z$ and effective mass $m^*$ are \emph{marginal} (should be
matched as accurately as possible), and all remaining
couplings---gradients, lattice nonlocality, residual vertex
structure---are \emph{irrelevant} at leading order.
This hierarchy is formulated near the Fermi surface.  The DiagMC
data discussed above (Fig.~\ref{fig:ueg_sigma}) show that it
extends to the entire occupied Fermi ball: $z(k)$ is nearly constant
and $m^*\approx m$, so a single matching at the Fermi surface fixes
the renormalization of all occupied states.  The Ward identity
$z\Gamma_3 = 1$ at $\bq = 0$ further ties the charge-vertex dressing to the
residue; DiagMC data~\cite{eft_sc} confirm that this cancellation
persists throughout the Fermi ball.
For the inhomogeneous valence liquid,
the locality of these properties means that the Fermi-surface
matching can be performed locally, at each point in real space:
this is what enables a local density approximation (LDA), matching
the RPT parameters to a local Fermi liquid at density $n(\br)$.

\emph{Relevant: density matching.}---The tree-level propagator
must reproduce $n(\br)$.  Because the xc interaction in the UEG
is short-ranged (screened), the static xc self-energy at any
point depends only on the local density---quantum Monte
Carlo~\cite{moroni1995,kukkonen2021} confirms this: the xc kernel
$f_{xc}(q) = d\Sigma_{xc}^{0}/dn$ (equivalently $F_0^s/N_0$ by
the compressibility sum rule) is nearly $q$-independent for
$q \lesssim 2k_F$.
Given this locality, the unique matching consistent with the
UEG limit fixes the static xc self-energy to
\begin{equation}\label{eq:static_xc_match}
    \Sigma_{xc}^{0}\bigl(n(\br)\bigr)
    \equiv
    \Sigma_{xc}^{\mathrm{UEG}}\bigl(n(\br);k_F,i0^+\bigr),
\end{equation}
a nonlinear functional of the local density.

\emph{Marginal: residue matching.}---The quasiparticle residue
$z_\Gamma^{\mathrm{val}}$ is fixed by the slope of the UEG
self-energy at the Fermi level.  Since integrating out the core
has already rescaled the frequency to
$\omega'=(z^{\mathrm{core}})^{-1}\omega$, and the approximate
Galilean invariance ensures $\Sigma$ depends
on $i\omega'$ and $\bk$ only through
$i\omega'-(\epsilon^{\mathrm{KS}}_{\nu\bk}-\epsilon_F)$, the
projection of the xc self-energy onto the matched subspace is
\begin{equation}\label{eq:Sigma_lin}
    P_2^\Gamma\Sigma^{\mathrm{val}}_{xc,\nu\bk}(i\omega')
    \equiv
    \Sigma_{xc}^{0}(n)
    + \bigl[1-(z_\Gamma^{\mathrm{val}})^{-1}\bigr]
    \bigl(i\omega'-(\epsilon^{\mathrm{KS}}_{\nu\bk}
    -\epsilon_F)\bigr),
\end{equation}
where $P_2^\Gamma$ extracts the relevant and marginal components
(Supplemental Material~\cite{supp}); $V_H$ is absorbed into
the propagator via the self-consistent density.

\emph{Tree-level Lagrangian.}---The matching conditions
    [Eqs.~\eqref{eq:static_xc_match} and~\eqref{eq:Sigma_lin}] complete
the RPT: the matched relevant and marginal couplings are absorbed
into the propagator as counterterms, leaving the residual
$\delta\Sigma = (1-P_2^\Gamma)\Sigma^{\mathrm{val}}$, which
contains only couplings that are irrelevant with respect to the
Fermi-liquid RG (self-energy curvature,
quasiparticle lifetime, and non-forward vertex corrections), as
the perturbation.  The tree level, $\delta\Sigma = 0$, gives
\begin{equation}\label{eq:L_KS}
    L_{\mathrm{KS}} =
    \bar\psi\,
    (z_\Gamma^{\mathrm{val}})^{-1}
    \Bigl[
        \hat Z_{\mathrm{core}}^{-1}\partial_\tau
        -\bigl(\hat H_{\mathrm{KS}}-\epsilon_F\bigr)
        \Bigr]\psi ,
\end{equation}
where
$(\hat Z_{\mathrm{core}}^{-1})_{\bG\bG'}
    =\delta_{\bG\bG'}+\sum_c f^c_{|\bk+\bG|}f^c_{|\bk+\bG'|}/\Delta E_c^2$
is the core renormalization operator (whose diagonal in the KS
eigenbasis gives $z^{\mathrm{core}}_{\nu\bk}$
[Eq.~\eqref{eq:zcore_nk}]; off-diagonal elements are suppressed
by $O(\delta^2/\Delta E_c^2) \sim 0.2\%$, see Supplemental
Material~\cite{supp}) and
$\hat H_{\mathrm{KS}}
    =-\nabla^2/2+V_{\mathrm{PSP}}+V_H[n]+\Sigma_{xc}^{0}(n)$,
where $\Sigma_{xc}^{0}$
[Eq.~\eqref{eq:static_xc_match}] is the matched static xc
self-energy---the LDA exchange-correlation potential
$V_{xc}^{\mathrm{LDA}}$.
Beyond tree level, $\delta\Sigma$ is expanded order by order
using KS propagators and subtracted vertices;
the Supplemental Material~\cite{supp} gives the Feynman rules and explicit
diagrams to second order.

Projecting Eq.~\eqref{eq:L_KS} onto a KS eigenstate gives
Eq.~\eqref{eq:G_factor}.  Since $\hat Z_{\mathrm{core}}^{-1}$
acts only on $\partial_\tau$ and not on $\hat H_{\mathrm{KS}}$,
the KS eigenvalue equation, self-consistent density, and total
energy are all unchanged from standard KS\@.  What changes is the
quasiparticle dispersion, which acquires the factor
$z^{\mathrm{core}}_{\nu\bk}$ [Eq.~\eqref{eq:qp_energy}].
For valence electrons, the Ward identity
$z^{\mathrm{val}}\Gamma_3^{\mathrm{val}}\approx1$ ensures that
$z^{\mathrm{val}}$ cancels in the pole position.
No such cancellation occurs for the core: the core polarizability
is suppressed by $\epsilon_F/\Delta E_c\sim0.02$--$0.05$, so
$\Gamma_3^{\mathrm{core}}\approx1$ rather than
$(z^{\mathrm{core}})^{-1}$, and $z^{\mathrm{core}}_{\nu\bk}$
survives in the quasiparticle energy.

\begin{figure}[t]
    \centering
    \includegraphics[width=0.95\columnwidth]{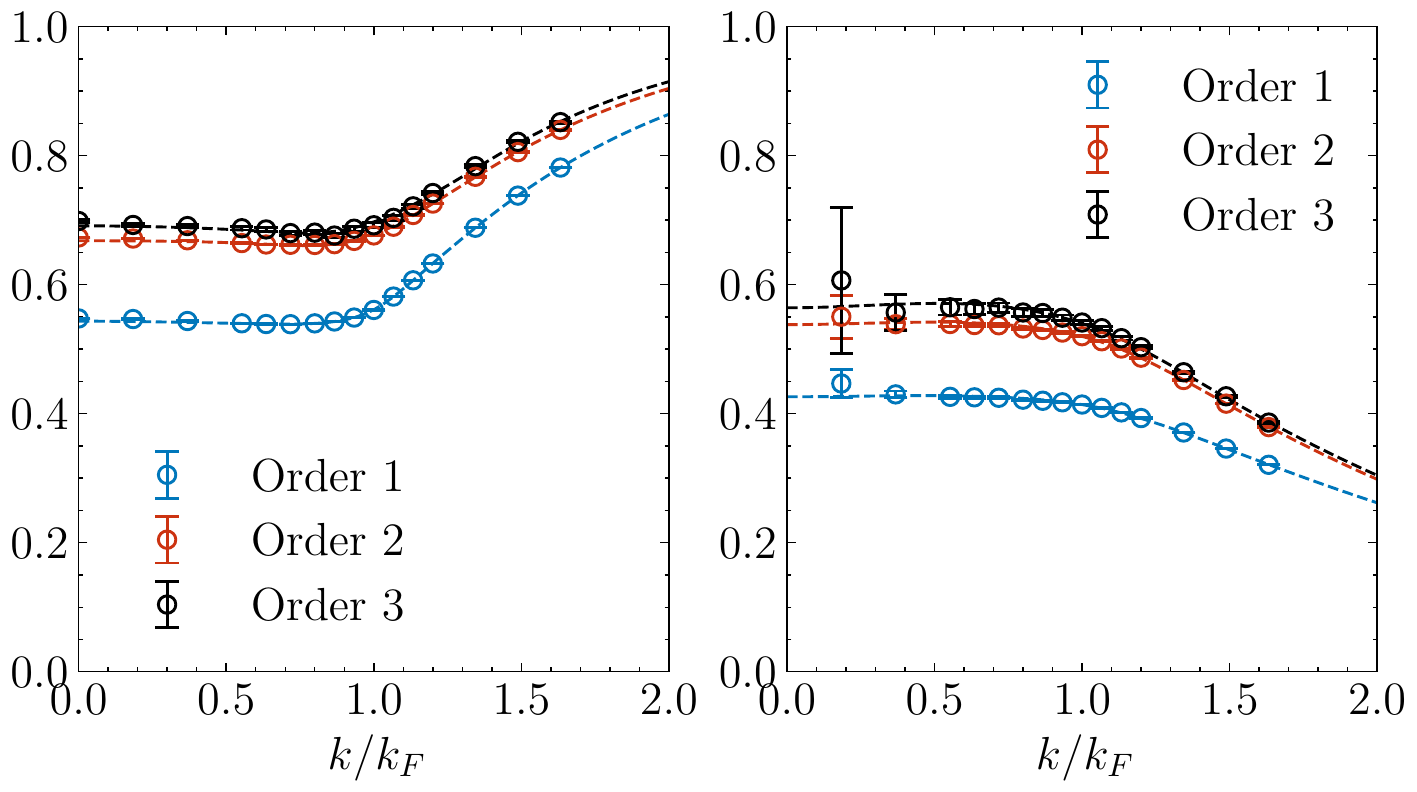}
    \caption{\label{fig:ueg_sigma}
        Quasiparticle properties of the UEG at $r_s = 4$ from
        DiagMC.
        Left: $z(k) = 1/(1 - \partial_\omega\mathrm{Re}\,\Sigma)$ is nearly
        constant within the Fermi sea.
        Right: $[\epsilon_k - \Delta\mathrm{Re}\,\Sigma(k)]/\epsilon_k$
        (where $\epsilon_k = k^2\!/2m$ and
        $\Delta\Sigma(k) = \Sigma(k) - \Sigma(0)$) is likewise flat,
        confirming $\partial_\omega\Sigma + \partial_{\epsilon}\Sigma
            \approx 0$ and hence $m^* \approx m$.}
\end{figure}

\textit{Results.---}We validate the theory across seven elements.
For lithium ($Z = 3$), the hydrogenic core
enables analytic evaluation of all integrals
(see Supplemental Material~\cite{supp}). Using a variational
Hartree--Fock 1s orbital, the static EFT potential agrees with
the conventional
GTH~\cite{goedecker1996} pseudopotential to $\sim$4\% at $\Gamma$,
confirming that the derivation recovers known pseudopotential
physics. This validates the two-step approach: a standard PSP
handles the static core screening, while the EFT provides the
dynamical correction.
For heavier elements (Na, K, Ca, Mg, Al, Si), core orbitals are
obtained numerically from a self-consistent radial KS solver
(see Supplemental Material~\cite{supp}).

The predicted QP band structures
(Fig.~\ref{fig:arpes_bands}, Table~\ref{tab:comparison})
confirm the central prediction: the correction is large
(20--35\%) for alkali and alkaline-earth metals, where
$\Delta E_c$ is only a few Hartree, and negligible ($<$5\%) for
Al and Si, where $\Delta E_c$ is several times larger.
The EFT agrees with eDMFT to better than
$0.15\;\eV$ for Li and K,
and reproduces the ARPES bandwidth for Na, Ca, Mg, and Al.
That a leading-order analytic formula matches a full quantum
Monte Carlo impurity calculation confirms that higher-order
corrections (suppressed by
$\epsilon_F/\Delta E_c \lesssim 0.05$) are small.
$G_0W_0$ narrows the bandwidth by only 4--10\%
(Table~\ref{tab:comparison}), confirming that perturbative
$GW$ captures valence screening but misses the core dynamical
effect.  PBE closely tracks LDA
(Fig.~\ref{fig:arpes_bands}), as detailed below.

The trend across elements is transparent: near $\Gamma$,
$1/z^{\mathrm{core}} - 1 \approx (f^c_0)^2/\Delta E_c^2$.
The form factor $f^c_0$ is of order the Coulomb self-energy $J_c$
of the core orbital ($\sim 1\;\Ha$, nearly $Z$-independent),
while $\Delta E_c$ grows rapidly with nuclear charge.
For alkali metals the outermost core $s$-orbital sits just below
the valence, giving $\Delta E_c$ of only a few Hartree and a
large correction.  Within the [Ne]-core series,
$z^{\mathrm{core}}_\Gamma$ increases monotonically
(Na 0.80, Mg 0.91, Al 0.96, Si 0.98) as $\Delta E_{2s}$ roughly
triples while $f^{2s}_0$ stays near $1\;\Ha$.
The non-monotonic ordering among alkali metals
(K $<$ Li $<$ Na) reflects different core shells:
K's $3s$ has the smallest $\Delta E_c$ ($1.72\;\Ha$).
For each atom, a single core $s$-orbital contributes
$>$99\% of the correction; deep core orbitals are suppressed by
their large $\Delta E_c$ ($\propto Z^2$), and $p$-orbitals vanish
near $\Gamma$ by angular momentum selection
($f_p \propto j_1 \to 0$).

\begin{table}[b]
    \caption{\label{tab:comparison}
        Occupied bandwidth ($\Gamma$-point depth in eV, measured below
        $\epsilon_F$ for metals and below VBM for Si).
        LDA and EFT QP: this work.
        $G_0W_0$ and eDMFT for Li--Mg from Ref.~\cite{mandal2022}.
        ARPES: Na~\cite{lyo1988,potorochin2022}; K~\cite{itchkawitz1990};
        Ca~\cite{sashin2000}; Mg~\cite{bartynski1986};
        Al~\cite{levinson1983}; Si~\cite{grobman1972}.}
    \begin{ruledtabular}
        \begin{tabular}{ldddddc}
               & \multicolumn{1}{c}{LDA}
               & \multicolumn{1}{c}{$G_0W_0$}
               & \multicolumn{1}{c}{eDMFT}
               & \multicolumn{1}{c}{EFT QP}
               & \multicolumn{1}{c}{EFT QP}
               & Expt                                                                                                                             \\
               &                              &                         &                         & \multicolumn{1}{c}{(LDA)}
               & \multicolumn{1}{c}{(PBE)}
               &                                                                                                                                  \\
            \hline
            Li & 3.48                         & 3.39                    & 2.60                    & 2.62                      & 2.55 & ---        \\
            Na & 3.27                         & 3.15                    & 2.84                    & 2.62                      & 2.63 & 2.65--2.78 \\
            K  & 2.27                         & 2.00                    & 1.42                    & 1.49                      & 1.44 & 1.60       \\
            Ca & 3.70                         & 3.79                    & 3.24                    & 3.06                      & 3.19 & 3.30       \\
            Mg & 6.92                         & 6.66                    & 6.18                    & 6.31                      & 6.34 & 6.15       \\
            \hline
            Al & 11.18                        & \multicolumn{1}{c}{---} & \multicolumn{1}{c}{---}
               & 10.70                        & 10.72                   & 10.6                                                                    \\
            Si & 12.26                        & \multicolumn{1}{c}{---} & \multicolumn{1}{c}{---}
               & 11.81                        & 11.85                   & 12.4                                                                    \\
        \end{tabular}
    \end{ruledtabular}
\end{table}

The correction is robust to the choice of both exchange-correlation
functional and pseudopotential.  Replacing the LDA functional and PSP
with PBE (functional and PSP together) at the same $Z_{\text{val}}$
changes the bandwidth narrowing by $< 0.2$ percentage points for Mg
and Al (Table~S5 and Fig.~S2 of the Supplemental
Material~\cite{supp}).
Even when $Z_{\text{val}}$ increases from $1$ to $9$ (Na, K) or $2$
to $10$ (Ca), promoting semicore shells into the valence, the
narrowing of the conduction band (the only band corrected, to avoid
double counting with the now-explicit semicore states) changes by
only $1$--$2$ percentage points.  This insensitivity is expected:
$z^{\mathrm{core}}_{\nu\bk}$ depends on core orbital properties
(form factors and excitation energies) computed from the isolated
atom, not from the solid-state functional or pseudopotential.

\textit{Discussion.---}The central message of this Letter is that
the Kohn--Sham Hamiltonian, while not derivable from DFT, emerges
as the tree-level output of the EFT of the electron gas.
The approximate Galilean invariance of the UEG and the
core--valence scale separation together force the tree-level
quasiparticle dispersion to coincide with KS eigenvalues,
up to the controlled correction $z^{\mathrm{core}}_{\nu\bk}$.
This reframes the KS Hamiltonian as the leading-order output of a
systematic many-body theory with a known expansion parameter
$\epsilon_F/\Delta E_c$, rather than an auxiliary construction
of DFT\@.

The EFT-KS framework also offers a natural route to resolving the
double-counting problem that plagues beyond-DFT methods.  In DFT+$GW$, the subtraction
$\Sigma_{GW} - V_{xc}$ is ad hoc because DFT provides no
systematic relationship between $V_{xc}$ and the self-energy
diagrams.  In DFT+DMFT, the double-counting correction
$\Sigma_{\mathrm{dc}}$ is even more ambiguous, with competing
prescriptions that lack a controlled justification.  In the EFT, both share the same language: the
RPT defines the tree-level propagator
(KS) and the residual $\delta\Sigma = (1-P_2^\Gamma)\Sigma$
through the same projector, so what is already included at tree
level and what remains as a correction is unambiguous at each order.

In this work we have applied the EFT to quasiparticle band
narrowing, the simplest observable sensitive to
$z^{\mathrm{core}}$.  The framework extends naturally in several
directions.  For semiconductors, the tree-level action retains an
unscreened long-range Coulomb component that, once included, yields
a KS Hamiltonian with hybrid-functional form, potentially providing
a first-principles explanation of the band-gap problem without
tuning parameters or $GW$-type diagrammatic calculations.
For strongly correlated
materials with $d$- or $f$-shell cores, where multiple dynamical
channels at comparable energies produce $l$-dependent quasiparticle
renormalization, the EFT-KS framework may provide a natural
embedding in which the impurity problem is formulated in the same
field-theoretic language as the lattice theory, potentially
eliminating the double-counting ambiguity that has long plagued
DFT+DMFT.
The methodology that produced this framework---first-principles
agentic science, in which LLM-based agents help reconstruct
theoretical foundations and then scale the verified theory as a
deterministic harness---is discussed in the End Matter, along with
other foundational problems across physics to which it may be
applied.

\begin{acknowledgments}
    \textit{Acknowledgments.---}We thank Gabriel Kotliar, Weiyi Guo, Zhiyi Li, Xinyang Dong, and Youjin Deng for
    discussions.
    K.C., X.C., and H.W.\ are supported by the National
    Natural Science Foundation of China under Grants
    No.~12474245 and No.~12447103, the National Key Research and
    Development Program of China under Grant No.~2024YFA1408604, and
    the GHfund~A (202407010637).
\end{acknowledgments}

\bibliography{refs}

% End Matter (PRL NEW 7/19/24 format)
% prl_endmatter.tex — PRL End Matter (NEW 7/19/24 PRL style)
% Included by prl_main.tex after \bibliography{refs}

\bigskip
{\centering\textbf{End Matter: First-principles Agentic Science}\par}
\medskip

\noindent AGI for Science~\cite{cai2025lac} promises a paradigm
shift---from human researchers assisted by computational tools to
LLM-based agents that participate substantively in theory
construction and validation.  Yet how to deploy such agents on
frontier scientific problems remains an open problem.  This Letter
proposes and demonstrates a concrete realization, which we call
\emph{first-principles agentic science}: LLM-based agents help
reconstruct a theoretical foundation with controlled accuracy, and
the verified theory then serves as a deterministic harness for
agentic scale-out.

Why is such a direction necessary?  AI for Science has achieved
impressive results \emph{within} established theoretical
frameworks: neural-network exchange--correlation
functionals~\cite{kirkpatrick2021} have solved the
fractional-electron problem; evolutionary symbolic
search~\cite{ma2022} has discovered new functional forms.  But
some problems are invisible from within the framework.  The
frozen-core renormalization $z^{\mathrm{core}}$ reported in this
Letter is a case in point: the ARPES bandwidth discrepancy
originates from dynamical core excitations that lie outside any
exchange--correlation functional, so no amount of fitting or
symbolic search within the Kohn--Sham framework can capture it.
Problems of this kind---where the theoretical framework itself
must be reconstructed---define the frontier that AI for Science
cannot reach and AGI for Science must address.

Why is it feasible?  Using LLM-based agents for foundational
reconstruction faces a hard verification challenge: at the
frontier no one knows the correct answer, reasoning chains are
long and fragile across multiple energy scales, and every output
must be audited by a human expert, so the verification cost grows
as $O(N_{\mathrm{applications}})$.  First-principles theoretical
problems have three remarkable features that resolve this tension.
First, no matter how long the reasoning chain, it only needs to be
verified once---at the symbolic level, against the internal
consistency of the first-principles framework.
Second, because the theory rests on controlled approximations, the
existing experimental record directly becomes an end-to-end
verification benchmark---no new experiments or labeled datasets
are needed.  Third, because the resulting theory is parameter-free,
it generalizes across systems by construction; a single verified
derivation therefore becomes a deterministic harness that agents
can scale to new materials indefinitely, with their role reduced
to tool invocation rather than open-ended reasoning---collapsing
the verification cost from $O(N_{\mathrm{applications}})$ to
$O(1)$.  In this way, the central difficulties of AI for
Science---verification, data sourcing, and
generalization---are structurally resolved.

This Letter instantiates all three features.  For the first, the
authors and agents jointly constructed an effective field theory of
the inhomogeneous electron liquid---a highly nontrivial task
requiring a single theory to unify the physics across three
disparate energy scales (core-electron excitations at several
Hartrees, the valence Fermi energy at a fraction of a Hartree, and
the low-energy excitations living immediately around the Fermi
surface), going beyond conventional EFT constructions due to the
intrinsic nonlocality of the extended Fermi surface; this
derivation was verified once at the symbolic level.  For the second, existing ARPES data for Li, Na, K, Ca, Mg, Al,
and Si directly provided both the benchmark and the end-to-end
verification---no new experiments or labeled datasets were needed,
and the parameter-free predictions passed without adjustment.  For
the third, the verified theory was implemented as a generic,
material-independent DFTK-based~\cite{dftk} pipeline; the agents
have already scaled it across the seven elements reported here,
and extension to further metals and compounds is underway.

Looking ahead, first-principles agentic science is not limited to
the present application.  Any first-principles theoretical
breakthrough shares the same three features---one-time
verifiability, falsifiability against existing data, and built-in
generalization---and therefore constitutes a high-leverage target
for agentic science at scale.  Across physics, open foundational
questions coexist with abundant experimental data waiting to be
confronted by controlled, parameter-free predictions: the
double-counting problem in strongly correlated $d$- and
$f$-electron materials, the pairing mechanism of unconventional
superconductors, the closure problem in turbulence, nuclear forces
derived from quantum chromodynamics, and the equation of state of
neutron stars constrained by gravitational-wave observations.
Each shares the structural prerequisites identified here, and each
could become the next target for first-principles agentic science.

% ═══════════════════════════════════════════
% Supplemental Material (appended)
% ═══════════════════════════════════════════
\clearpage
\onecolumngrid
\begin{center}
\textbf{\large Supplemental Material}
\end{center}
\twocolumngrid

\setcounter{secnumdepth}{3}
\renewcommand{\thesection}{\Roman{section}}

%=============================================================================
\section{Dual-fermion derivation of the pseudopotential}
\label{app:derivation}
%=============================================================================

We derive the valence pseudopotential $\delta V_{pp}$ by integrating
out core electrons from the Born--Oppenheimer action using the
dual-fermion method of Rubtsov et al.~\cite{rubtsov2008,rubtsov2012,ribic2017}.  The
result is that the valence propagator takes the form
$g_v^{-1} = g_0^{-1} + \delta V_{pp}$, where $\delta V_{pp}$
is the energy-dependent pseudopotential whose matrix elements are
derived explicitly in Sec.~\ref{app:lambda_expansion}.

The starting point is the Born--Oppenheimer Lagrangian in Hartree
atomic units:
\begin{align}\label{eq:lagrangian}
    L &= \int_{\br\sigma} \barpsi_{\br\sigma}\!
      \left[\partial_\tau - \frac{\nabla^2}{2}
      + V_{\mathrm{Lat}} - \mu\right]\!
      \psi_{\br\sigma} \notag\\
    &\quad + \frac{1}{2}\int_{\br\br'}
      \frac{\bar\psi_{\br\sigma}\bar\psi_{\br'\sigma'}
      \psi_{\br'\sigma'}\psi_{\br\sigma}}{|\br-\br'|},
\end{align}
where $V_{\mathrm{Lat}}$ is the lattice (electron--nuclear) potential
and $\mu$ the chemical potential.
We introduce a reference system $g_c$ that captures the core by
adding a hybridization $R$ to the non-interacting propagator
$g_0^{-1} = -i\omega - \nabla^2/2 + V_{\mathrm{Lat}} - \mu$:
\begin{equation}\label{eq:gc}
    g_c^{-1} = g_0^{-1} + R.
\end{equation}
The hybridization $R$ serves two purposes: (i)~it pushes valence
orbitals to energy $\Lambda \to \infty$, freezing them in the
reference, and (ii)~it ensures that the core sector of $g_c$
reproduces the physical core propagator, including
electron--electron screening among core electrons.  Since $g_0$
contains only the bare nuclear potential, $R$ must absorb the
core mean field; in practice, the Hartree--Fock self-energy of
the core provides an excellent approximation for closed-shell
cores.

To integrate out the core, we decouple $R$ via a
Hubbard--Stratonovich transformation, introducing auxiliary
fields $\phi$:
\begin{equation}
    e^{\int_{XX'} \bar\psi_X R \psi_{X'}}
    = \frac{1}{\det R^{-1}}
    \int \calD[\bar\phi,\phi]\,
    e^{-\bar\phi R^{-1} \phi - (\bar\phi\psi + \bar\psi\phi)}.
\end{equation}
The partition function becomes
$Z \propto \int \calD[\bar\phi,\phi]\,
e^{-\bar\phi R^{-1}\phi}
\int \calD[\bar\psi,\psi]\,
e^{-S_c[\bar\psi,\psi] - (\bar\phi\psi + \bar\psi\phi)}$,
where $S_c$ is the reference action with $g_c$ as the bare
propagator.  Integrating out $\psi$ in the presence of $S_c$
produces the cumulant expansion
$\int \calD[\bar\psi,\psi]\, e^{-S_c - (\bar\phi\psi + \bar\psi\phi)}
= e^{\bar\phi\, G_2\, \phi + O(\phi^4)}$,
where $G_2$ is the interacting two-point function of the reference
system.  Truncating at the two-point level (the $O(\phi^4)$ terms
generate valence--valence interactions already present in $L$) and
rescaling $\phi \to g_c^{-1}\psi$ to restore the original field
normalization, the effective valence propagator is:
\begin{equation}\label{eq:gv}
    g_v^{-1} = g_c^{-1}\!\left[R^{-1} - G_2\right]\!g_c^{-1}.
\end{equation}
Substituting $g_c^{-1} = g_0^{-1} + R$ and expanding:
\begin{align}
    g_v^{-1} &= (g_0^{-1} + R)\, R^{-1}\, (g_0^{-1} + R)
    - g_c^{-1} G_2\, g_c^{-1}.
\end{align}
Using $G_2^{-1} = g_c^{-1} - \Sigma_c$, the Dyson equation for
the reference system ($\Sigma_c$ is the self-energy from
Coulomb interactions within the reference), and simplifying
yields:
\begin{equation}\label{eq:gv_result}
    \boxed{g_v^{-1} = g_0^{-1} + \delta V_{pp}},
\end{equation}
with the pseudopotential
\begin{equation}\label{eq:dvpp_def}
    \delta V_{pp} = g_0^{-1}\, R^{-1}\, g_0^{-1}
    + \Sigma_c
    + \Sigma_c\, G_2\, \Sigma_c\,.
\end{equation}
The first term diverges for core matrix elements as
$\Lambda \to \infty$, projecting out core states.  The total
$\delta V_{pp}$ in the valence sector carries the physical
core--valence interaction: Hartree and exchange screening of the
nucleus, core--valence orthogonality, and dynamical poles at the
core excitation energies $\Delta E_c$.  How these effects are
distributed among the three terms depends on the choice of $R$;
the explicit matrix elements are obtained via the $1/\Lambda$
expansion in the next section.

%=============================================================================
\section{$1/\Lambda$ expansion and matrix elements}
\label{app:lambda_expansion}
%=============================================================================

This section evaluates the abstract $\delta V_{pp}$ of
Eq.~\eqref{eq:dvpp_def} by expanding the interacting Green's function
$G_2$ of the \emph{reference system} in powers of $1/\Lambda$ and
taking $\Lambda \to \infty$.  The result is the pseudopotential
formula [Eq.~(6) of the main text] with explicit Hartree--Fock and
dynamical matrix elements.

We assume core electrons are localized on individual ions.  The
reference system (Sec.~\ref{app:derivation}) is an interacting
atom in which valence orbitals are pushed to energy
$E_0 + \Lambda$ ($\Lambda \to \infty$) while core orbitals remain
at their physical energies.  As discussed above, $R$ includes the
core mean field, so the single-particle core orbitals $\{j\}$
entering $g_c$ are at the Hartree--Fock level (atomic Kohn--Sham
orbitals for Na--Si, or the variational single-exponential form
for Li; see Sec.~\ref{app:numerical}).
The reference Hamiltonian in this basis is
\begin{equation}\label{eq:Hc}
    \hat{H}_c = \sum_j E_j^c\, a_j^\dagger a_j
    + \frac{1}{2}\sum_{ijkl} v_{ijkl}\, a_i^\dagger a_j^\dagger a_l a_k,
\end{equation}
where $E_j^c = E_j$ (the orbital energy from $g_c$) for core
orbitals, $E_j^c = E_0 + \Lambda$ for valence orbitals, and
$v_{ijkl} = \eint{ij}{kl}$ are antisymmetrized two-electron
integrals ($\eint{ij}{kl} \equiv \langle ij|kl\rangle - \langle ij|lk\rangle$).

The ground state of this reference system is
\begin{equation}\label{eq:gs}
    \ket{\Phi_0} = \ket{N_c,\xi_0}
    + \sum_{\xi_1}
    \frac{\bra{N_c,\xi_1}\hat V_{ee}\ket{N_c,\xi_0}}{n(\xi_1)\Lambda}
    \ket{N_c,\xi_1} + O(1/\Lambda^2),
\end{equation}
where $\hat V_{ee}$ is the Coulomb interaction,
$\ket{N_c,\xi_0}$ is the Slater determinant with all $N_c$
core orbitals filled (no valence electrons),
$\ket{N_c,\xi_1}$ denotes configurations with one valence orbital
occupied, and $n(\xi_1)$ counts the number of valence orbitals in
the configuration.

Labeling core orbitals as $i,j$ and
valence orbitals as $s,t$, the interacting Green's function of the
reference system decomposes as
$(G_2)_{st} = (G_2^p)_{st} + (G_2^h)_{st}$.

\textit{Particle part.}---Expanding $N_c+1$-particle states to
$O(1/\Lambda^2)$:
\begin{equation}
    (G_2^p)_{st} = \frac{\delta_{st}}{\Lambda}
    + \frac{i\omega\,\delta_{st} - \Sigma^{\text{HF}}_{st}}{\Lambda^2}
    + O(1/\Lambda^3),
\end{equation}
where $\Sigma^{\text{HF}}_{st} = \sum_c n_c\,
[\langle cs|ct\rangle - \langle cs|tc\rangle]$,
with $\langle cs|ct\rangle$ the direct (Hartree) and
$\langle cs|tc\rangle$ the exchange two-electron integral.

\textit{Hole part.}---The ground state has $O(1/\Lambda)$ admixture:
\begin{equation}
    (G_2^h)_{st} = -\frac{1}{\Lambda^2}\sum_{c}
    \frac{M_{sc}\, M_{tc}^*}{i\omega + \Delta E_c},
\end{equation}
where $\Delta E_c = E_c^{(\mathrm{hole})} - E_0$ is the energy to
create a hole in core orbital $c$ ($E_0$ is the $N_c$-electron
ground state energy) and
$M_{sc} = \bra{N_c,\xi_1^{sc}}\hat V_{ee}\ket{N_c,\xi_0}$ is the
Coulomb matrix element for the core-to-valence excitation
$c \to s$.

Combining and taking $\Lambda \to \infty$:
\begin{equation}
    (\delta V_{pp})_{st} = \Sigma^{\text{HF}}_{st}
    + \sum_c \frac{M_{sc}\, M_{tc}^*}{i\omega + \Delta E_c},
\end{equation}
which is the pseudopotential formula [Eq.~(6) of the main text]
in the valence sector.  Core--core matrix elements of
$g_v^{-1}$ diverge as $\Lambda \to \infty$, projecting out
core states.

\subsection{Four-term formula in coordinate space}
\label{app:four_terms}

Evaluating this in coordinate space for a core of
spin-paired orbitals $\{\phi_c\}$ with radial functions $u_c(r) = r R_c(r)$:
\begin{widetext}
\begin{equation}\label{eq:dvpp_ion}
\begin{aligned}
    \delta V_{pp}^{(\text{ion})}(\br,\br')
    &= \underbrace{2\sum_c V_{H,c}(r)\,\delta(\br - \br')}_{\text{Term 1: Hartree}}
       \underbrace{- \sum_c \phi_c^*(\br)\,\phi_c(\br')\,
       \frac{1}{|\br - \br'|}}_{\text{Term 2: Exchange}} \\
    &\quad \underbrace{- \sum_c \phi_c^*(\br)\,\phi_c(\br')\,
             \bigl[V_{H,c}(r) + V_{H,c}(r') - 2J_c\bigr]}_{\text{Term 3: Projection (static $\Sigma G_2 \Sigma$)}}
       \underbrace{+ \sum_c \phi_c^*(\br)\,\phi_c(\br')\,
             \frac{\bigl(V_{H,c}(r) - J_c\bigr)\bigl(V_{H,c}(r') - J_c\bigr)}
             {i\omega + \Delta E_c}}_{\text{Term 4: Dynamical}},
\end{aligned}
\end{equation}
\end{widetext}
where $c$ sums over spatial core orbitals (spin-paired),
the factor 2 in Term~1 counts both spins (Hartree is
spin-independent), while Terms~2--4 carry no such factor
(exchange and projection act between same-spin electrons only),
$V_{H,c}(r) = r^{-1}\int_0^r u_c^2\,dr' + \int_r^\infty u_c^2/r'\,dr'$
is the Hartree potential of orbital $c$,
$J_c = \int u_c^2\, V_{H,c}\,dr$ is the self-Coulomb integral,
and $\Delta E_c = E_{\text{core}}^{(\text{hole}\,c)} -
E_{\text{core}}^{(\text{gs})}$ is the core excitation energy.

The first three terms are frequency-independent and define a
conventional (static) pseudopotential.  Term~4 depends on
frequency $\omega$ through the core excitation poles at
$\Delta E_c$.  Its value at $\omega = \epsilon_F$ is a static
shift already absorbed into fitted PSP parameters (see the
double-counting subtraction in Sec.~\ref{app:dyson}); only
the frequency \emph{dependence} around $\epsilon_F$ is new.
Because
$\partial_\omega V_{\mathrm{dyn}}|_{\epsilon_F} \neq 0$,
the quasiparticle residue acquires a frozen-core renormalization
factor that we call $z^{\mathrm{core}}$.  For elements with
low-lying core excitations ($\Delta E_c \sim 2$--$3\;\Ha$), this
correction reduces the quasiparticle bandwidth by up to
$\sim$35\%.  The closed-form expression for
$z^{\mathrm{core}}_{\nu\bk}$ and its derivation from the Dyson
equation are given in Sec.~\ref{app:dyson}.

\paragraph{Kleinman--Bylander projectors and vertex cancellation.}
In practice, conventional pseudopotentials use the
Kleinman--Bylander (KB) form with nonlocal projectors.  The
Ward identity $z^{\mathrm{val}}\Gamma_3 \approx 1$ ensures that
the static potentials ($V_{\mathrm{PSP}}$, $V_H$) enter the
KS Hamiltonian without an additional valence residue for local
(multiplicative) potentials.  For nonlocal KB projectors, the
formal vertex residual from the $z\Gamma_3$ cancellation is
$l$-dependent; however, it is absorbed by the fitted KB projector
parameters ($h_{ij}^l$ coefficients), which are determined
empirically to reproduce atomic scattering properties.  The
EFT therefore inherits the accuracy of the underlying KB
pseudopotential for the static sector, with $z^{\mathrm{core}}$
providing the dynamical correction on top.

%=============================================================================
\section{Renormalized perturbation theory}
\label{app:rpt}
%=============================================================================

\subsection{Formal setup}

We organize the valence many-body problem as a renormalized
perturbation theory (RPT): the KS Hamiltonian defines the
zeroth-order propagator, and corrections are expanded in
powers of the valence Coulomb interaction with matched
contributions subtracted at each order.

The Hartree potential $V_H[n]$ is determined self-consistently
by the SCF density and enters $H_{\mathrm{KS}}$ undressed
($z^{\mathrm{val}}\Gamma_3 \approx 1$ to $\sim$3--5\%).
We therefore absorb $V_H$ into $g^{\mathrm{KS}}$ and expand
only the exchange-correlation self-energy
$\Sigma^{\mathrm{val}}_{xc}$ diagrammatically.

The tree-level propagator in the KS eigenbasis is
\begin{equation}\label{eq:gks_diag_app}
    g^{\mathrm{KS}}_{\nu}(\bk,i\omega_n)
    =
    \frac{\mathcal{R}_{\nu\bk}}{i\omega_n-E_{\nu\bk}^{(0)}}\,,
    \quad
    E_{\nu\bk}^{(0)}
    =
    z^{\mathrm{core}}_{\nu\bk}
    (\epsilon^{\mathrm{KS}}_{\nu\bk}-\epsilon_F),
\end{equation}
with residue $\mathcal{R}_{\nu\bk}=z^{\mathrm{val}}z^{\mathrm{core}}_{\nu\bk}$.
The frozen-core factor $z^{\mathrm{core}}_{\nu\bk}$
(Sec.~\ref{app:lambda_expansion}) enters because $g^{\mathrm{KS}}$
includes the full pseudopotential; the dynamical part
(Term~4) rescales the pole position relative to $\epsilon_F$.
The valence weight $z^{\mathrm{val}}$ arises from
valence--valence interactions and is determined
self-consistently by the RPT matching conditions below.
Setting $\delta\Sigma = 0$ recovers the EFT quasiparticle
energies $\epsilon_F + z^{\mathrm{core}}(\epsilon^{\mathrm{KS}}
- \epsilon_F)$ as pole positions;
$z^{\mathrm{val}}$ multiplies the residue but does not shift
the poles.

Two projectors define the matching.  Working in the plane-wave
basis (reciprocal lattice vectors $\bG$, $\bG'$),
$P_2^\Gamma$ extracts the local ($\bq = 0$) xc potential and the
$\Gamma$-point frequency slope from the xc self-energy,
\begin{align}\label{eq:p2_sigma_def}
    &(P_2^\Gamma\Sigma_{xc})_{\bG\bG'}(\bk,i\omega)
    =
    \bigl[\Sigma_{xc}^{0}(n)\bigr]_{\bG-\bG'} \notag\\
    &\quad +
    \bigl(1-(z_\Gamma^{\mathrm{val}})^{-1}\bigr)
    \bigl[i\omega\,\delta_{\bG\bG'}
    -(H_{\mathrm{KS}}-\epsilon_F)_{\bG\bG'}\bigr],
\end{align}
where
$\Sigma_{xc}^{0}(n) = \Sigma_{xc}^{\mathrm{UEG}}(n;k_F,i0^+)$
is the exchange-correlation potential determined by the
density-response matching condition (see main text),
$(H_{\mathrm{KS}})_{\bG\bG'}$ the KS Hamiltonian matrix,
and $z_\Gamma^{\mathrm{val}}$ the valence $z$-factor evaluated at
the $\Gamma$ point.
The use of $H_{\mathrm{KS}} - \epsilon_F$ (rather than the
free-particle dispersion $\xi_{\bk+\bG}$) ensures that the
$z$-rescaling applies to all static potentials, not just the
kinetic energy; physically, this reflects the approximate
Galilean invariance of the KS band structure.
$P_3^0$ extracts the s-wave limit of the charge vertex
$\Gamma_3$ (the dressed density--density coupling) on the Fermi
surface,
\begin{equation}\label{eq:p3_gamma_def}
    P_3^0\Gamma_3
    =
    \left\langle
    \lim_{\bq\to0}\lim_{\Omega\to0}
    \Gamma_3(\bk,i0^+;\bq,i\Omega)
    \right\rangle_{\mathrm{FS}},
\end{equation}
with renormalization condition
$z_\Gamma^{\mathrm{val}}\,P_3^0\Gamma_3^{\mathrm{val}}=1$.
The two counterterms
\begin{equation}\label{eq:p2p3_def}
    \mathcal C_2^{(n)}
    =
    - P_2^\Gamma\Sigma_{\mathrm{eff}}^{(n)},\quad
    \mathcal C_3^{(n)}
    =
    - P_3^0\Gamma_{3,\mathrm{eff}}^{(n)}
\end{equation}
subtract the matched contributions at each order $n$, where
$\Sigma_{\mathrm{eff}}^{(n)}$ includes all order-$n$ self-energy
diagrams built from $v^{\mathrm{val}}$ vertices,
$g^{\mathrm{KS}}$ lines, and lower-order counterterm insertions.
The residual at each order is
$\delta\Sigma^{(n)} = (1-P_2^\Gamma)\Sigma_{\mathrm{eff}}^{(n)}$.

\subsection{Why tree level captures the dominant physics}

Tree level ($\delta\Sigma = 0$) gives the KS Hamiltonian by
construction.  The non-trivial claim is that $\delta\Sigma$ is
\emph{small}.  We now make this explicit.

The valence Dyson equation reads
$g^{-1} = g_v^{-1} + V_H[n] + \Sigma^{\mathrm{val}}_{xc}$,
where $g_v$ is the non-interacting valence
propagator [Eq.~\eqref{eq:gv_result}] and
$\Sigma^{\mathrm{val}}_{xc}$ is the xc self-energy
(plus sign as in $g_v^{-1} = g_0^{-1} + \delta V_{pp}$).
Since $V_H$ is absorbed into $g^{\mathrm{KS}}$ via the SCF,
the xc counterterm is
$\Sigma_{\mathrm{CT}} \equiv (g^{\mathrm{KS}})^{-1}
- g_v^{-1} - V_H$, giving
$g^{-1} = (g^{\mathrm{KS}})^{-1} + \delta\Sigma$ with
\begin{equation}\label{eq:deltaSigma_explicit}
  \boxed{\delta\Sigma
  = \Sigma^{\mathrm{val}}_{xc} - \Sigma_{\mathrm{CT}}}\,.
\end{equation}
The counterterm is:
\begin{equation}\label{eq:Sigma_CT}
  \Sigma_{\mathrm{CT}}
  = V_{xc}^{\mathrm{LDA}}[n]
  + \bigl(1 - (z_\Gamma^{\mathrm{val}})^{-1}\bigr)
  \bigl(i\omega' - (\epsilon^{\mathrm{KS}}_{\nu\bk}
  - \epsilon_F)\bigr),
\end{equation}
where $\omega' = (z^{\mathrm{core}})^{-1}\omega$ is the
renormalized frequency and
$\epsilon^{\mathrm{KS}}_{\nu\bk} - \epsilon_F$ the KS
eigenvalue relative to the Fermi level.
The claim ``tree level is good'' is the claim that
$\Sigma_{\mathrm{CT}}$ captures the dominant part of the
\emph{unknown} $\Sigma^{\mathrm{val}}_{xc}$, so that their
difference $\delta\Sigma$ is small.

The projector $P_2^\Gamma$ makes this manifest:
$P_2^\Gamma\Sigma^{\mathrm{val}}_{xc} = \Sigma_{\mathrm{CT}}$
by construction, so
$\delta\Sigma = (1 - P_2^\Gamma)\Sigma^{\mathrm{val}}_{xc}$.
Each component of $\Sigma_{\mathrm{CT}}$ cancels a corresponding
large piece of $\Sigma^{\mathrm{val}}_{xc}$:

\textit{Exchange-correlation ($V_{xc}^{\mathrm{LDA}}$)}: cancels
the local-density part of the xc self-energy.  Residual:
the nonlocal xc gradient, controlled by
$f_{xc}(q) - f_{xc}(0)$, where $f_{xc}$ is the static
exchange-correlation kernel defined by
$\chi_0^{-1} = \chi^{-1} + v_c + f_{xc}$
($\chi_0$: Lindhard function, $\chi$: interacting
susceptibility).
Quantum Monte Carlo calculations confirm that $f_{xc}(q) \approx
f_{xc}(0)$ for $q \lesssim 2k_F$~\cite{moroni1995,kukkonen2021},
with deviations of $\sim$20--30\% only at
$q \gtrsim 2.5\,k_F$.

\textit{Frequency slope ($z$-factor)}: the
$(1 - (z_\Gamma^{\mathrm{val}})^{-1})(i\omega' - \xi)$ term
cancels the leading frequency dependence of
$\Sigma^{\mathrm{val}}$ at the $\Gamma$ point.  Residual:
$((z_\Gamma^{\mathrm{val}})^{-1} - z^{\mathrm{val}}(k)^{-1})
(i\omega' - \xi)$, small because $z^{\mathrm{val}}(k)$ is nearly
constant within the Fermi sea
(Fig.~3 of the main text, left panel).

\textit{Effective mass}: the Galilean-invariant combination
$i\omega' - \xi_\bk$ absorbs the dominant dispersion but not
the mass renormalization; the residual
$\propto (m^*/m - 1) \sim 1$--$3\%$ is genuinely uncancelled
(Fig.~3 of the main text, right panel).

\textit{Vertex renormalization of $V_{\mathrm{PSP}}$.}---The
static pseudopotential $V_{\mathrm{PSP}}$
(Terms~1--3 of Sec.~\ref{app:four_terms}) couples to the
quasiparticle through the dressed charge vertex $\Gamma_3$.
The $P_3^0$ matching condition
$z_\Gamma^{\mathrm{val}}\,P_3^0\Gamma_3 = 1$
[Eq.~\eqref{eq:p3_gamma_def}] absorbs the $\bq = 0$ vertex
exactly, so that $z^{\mathrm{val}}\Gamma_3 = 1$ at forward
scattering and $V_{\mathrm{PSP}}$ enters $H_{\mathrm{KS}}$
undressed.  The same cancellation applies to $V_H$ (already
absorbed into $g^{\mathrm{KS}}$ via the SCF).
The finite-$\bq$ residual
$(1-P_3^0)(z^{\mathrm{val}}\Gamma_3 - 1)\,V$ is projected into
$\delta\Sigma$ as a higher-order correction;
DiagMC~\cite{eft_sc} shows this residual is $\sim$3--5\%
throughout the Fermi ball.

In total, $|\delta\Sigma|/\epsilon_F$ is bounded at a few
percent for the occupied bandwidth.

\subsection{Beyond tree level}

\begin{figure*}[t]
\centering
\includegraphics[width=0.95\textwidth]{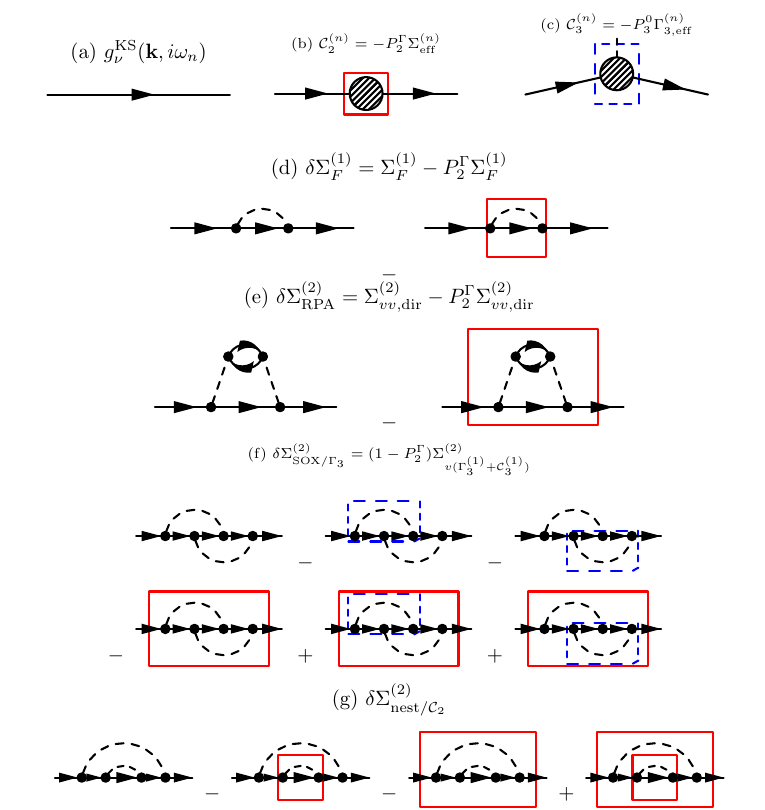}
\caption{\label{fig:rpt_diagrams}
RPT self-energy diagrams to second order.
(a) KS propagator. (b,c) Two- and three-point counterterms.
(d) First-order Fock residual.
(e) RPA/bubble. (f) SOX/charge-vertex. (g) Nested Fock.
Solid boxes (red): $P_2^\Gamma$ subtraction; dashed boxes (blue): $P_3^0$ subtraction.}
\end{figure*}

The residual $\delta\Sigma$ can be evaluated diagrammatically
order by order.  Figure~\ref{fig:rpt_diagrams} shows the RPT
diagrams to second order using standard Matsubara rules: solid
lines are $g^{\mathrm{KS}}$, dashed lines are the bare Coulomb
interaction $v^{\mathrm{val}}$, solid boxes (red) denote
$P_2^\Gamma$ subtraction, and dashed boxes (blue) denote $P_3^0$
subtraction.

At first order [panel~(d)], only the Fock (exchange) diagram
contributes (Hartree is in the propagator); the residual is
$\delta\Sigma^{(1)} = (1 - P_2^\Gamma)\Sigma_F^{(1)}$.
At second order [panels~(e)--(g)],
$\Sigma^{(2)}_{\mathrm{eff}}$ includes the bare two-Coulomb
diagrams (RPA/bubble, SOX, nested Fock) together with mixed
diagrams containing one $\mathcal C_2^{(1)}$ or
$\mathcal C_3^{(1)}$ counterterm insertion; the residual is
$\delta\Sigma^{(2)} = (1 - P_2^\Gamma)\Sigma^{(2)}_{\mathrm{eff}}$.

These diagrams are organized in powers of the bare Coulomb
interaction $v^{\mathrm{val}}$.  For practical calculations beyond
tree level, the expansion should be reorganized around the
RPA-screened interaction
$W_{\mathrm{KS}} = v/(1 - v\chi_0^{\mathrm{KS}})$, which resums
the screening chain and is short-ranged.  The matching conditions
$P_2^\Gamma$ and $P_3^0$ are unchanged.  In the $W$-expanded
form, the first-order correction is the standard $G_0W_0$
approximation (screened exchange $+$ Coulomb hole), connecting
the RPT to the one-loop correction referenced in the main text.
A formally exact expansion with order-by-order cancellation
(including Hartree diagrams, a screening projector, and
interaction counterterms) is straightforward but beyond the
scope of this work.

%=============================================================================
\section{Derivation of the quasiparticle formula}
\label{app:dyson}
%=============================================================================

We derive the closed-form expression for
$z^{\mathrm{core}}_{\nu\bk}$ and the QP energy formula
[Eq.~(2) of the main text] from the Dyson equation for the
separable $\Sigma_{\mathrm{dyn}}$ of
Sec.~\ref{app:lambda_expansion}, showing that the result is exact
up to corrections of order
$(\epsilon_{\text{KS}} - \epsilon_F)^2/\Delta E^2 \sim 0.2\%$.

\subsection{KS basis and self-energy projection}

Standard DFT diagonalizes the static KS Hamiltonian
$H_{\text{KS}} = T + V_{\text{stat}} + V_H + V_{\text{xc}}$,
where $V_{\text{stat}}$ is the frequency-independent part of the
pseudopotential (Terms~1--3 of Sec.~\ref{app:four_terms}),
giving eigenstates $\ket{\psi_{\nu\bk}}$ with eigenvalues
$\epsilon_{\nu\bk}$ and Bloch coefficients $c_{\nu\bk}(\bG)$.
Projecting the dynamical part $\Sigma_{\text{dyn}}$ (Term~4) onto
this basis, define the coherent form factor
\begin{equation}
  F^c_{\nu\bk} = \sum_{\bG} c_{\nu\bk}(\bG)\,f^c_{|\bk+\bG|},
\end{equation}
where $f^c_K$ is obtained by Fourier-transforming the core
coupling in Term~4 of Eq.~\eqref{eq:dvpp_ion}:
$f^c_K = \int d\br\, e^{-i\bK\cdot\br}\,
\phi_c^*(\br)\,(V_{H,c}(r) - J_c)$.
For s-wave core orbitals ($\phi_c = u_c(r)/(r\sqrt{4\pi})$),
the angular integral selects $l = 0$ and
$j_0(Kr) = \sin(Kr)/(Kr)$ yields the radial form
[Eq.~\eqref{eq:form_factor} of the main text]:
$f^c_K = (\sqrt{4\pi}/K)\int_0^\infty
u_c(r)\,[V_{H,c}(r) - J_c]\,\sin(Kr)\,dr$
(see Sec.~\ref{app:numerical} for the numerical quadrature).
The self-energy matrix in the KS basis is
\begin{equation}\label{eq:sigma_nn_dyson}
  \Sigma_{\nu\nu'}(\bk;\omega)
  = \sum_c \frac{F^c_\nu\,(F^c_{\nu'})^*}{\omega + \Delta E_c},
\end{equation}
which is \emph{not} diagonal in band index $\nu$.

\subsection{Double counting and subtraction}

The full Dyson equation
$\det[\omega\,\delta_{\nu\nu'} - \epsilon_{\nu\bk}\,\delta_{\nu\nu'}
- \Sigma_{\nu\nu'}(\bk;\omega)] = 0$
gives unphysical results because $\Sigma_{\text{dyn}}$ contains a
static piece.  For a single channel (dropping the sum over $c$
and band indices for clarity):
\begin{equation}
  \Sigma_{\text{dyn}}(\omega)
  = \underbrace{\frac{|f\rangle\langle f|}{\epsilon_F
  + \Delta E_c}}_{\displaystyle \Sigma_0}
  + \underbrace{\frac{|f\rangle\langle f|\,(\epsilon_F - \omega)}
  {(\omega+\Delta E_c)(\epsilon_F+\Delta E_c)}}
  _{\displaystyle \delta\Sigma(\omega)}.
\end{equation}
The static part $\Sigma_0 \sim 20\;\eV$ (for Li) is already included
in $H_{\text{KS}}$ via $V_{\text{xc}}$ and the pseudopotential.
Including it again double-counts.

The subtracted diagonal self-energy is
\begin{align}
  \Sigma_{\nu\nu}(\omega) - \Sigma_{\nu\nu}(\epsilon_F)
  &= -\frac{|F^c_\nu|^2\,(\omega - \epsilon_F)}
  {(\omega+\Delta E_c)(\epsilon_F+\Delta E_c)}.
\end{align}

\subsection{Exact solution and $z$-factor}

Setting $\delta = \omega - \epsilon_F$ and
$s = \epsilon_{\nu\bk} - \epsilon_F$, with
$D = \epsilon_F + \Delta E_c$:
\begin{equation}
  \delta = \frac{s\,D(\delta+D)}{D(\delta+D) + |F^c_\nu|^2}.
\end{equation}
Since $|\delta| \sim 0.1\;\Ha \ll D \sim 2.7\;\Ha$, setting
$\delta + D \approx D$ gives
\begin{equation}
  \delta \approx \frac{s}{1 + |F^c_\nu|^2/D^2}
  = z^{\mathrm{core}}_{\nu\bk}\,
  (\epsilon_{\nu\bk} - \epsilon_F),
\end{equation}
where $z^{\mathrm{core}}_{\nu\bk} = 1/(1 + |F^c_\nu|^2/D^2)
\approx 1/(1 + |F^c_\nu|^2/\Delta E_c^2)$,
recovering the QP energy formula [Eq.~(2) of the main text] with
$z^{\mathrm{core}}_{\nu\bk} = 1/(1 + \sum_c |F^c_{\nu\bk}|^2/\Delta E_c^2)$.
No linearization in $\omega$ was used; the only approximation
is $\delta/D \sim 4\%$.

\subsection{Off-diagonal (band mixing) is suppressed}

The subtracted off-diagonal matrix elements
$\Sigma_{\nu\nu'}^{\text{sub}}(\omega)
= -F^c_\nu (F^c_{\nu'})^*\,(\omega - \epsilon_F)/[(\omega+\Delta E_c)D]$
vanish as $\omega \to \epsilon_F$.
The second-order correction from band mixing is
$\delta\epsilon_\nu^{(2)} \sim |F^c_\nu|^2|F_{\nu'}|^2\,\delta^2
/ [D^4\,(\epsilon_\nu - \epsilon_{\nu'})]
= O(\delta^2/D^2) \sim 0.2\%$,
making the diagonal approximation excellent.

\section{Numerical results}
\label{app:numerical}
%=============================================================================

This section describes the computational pipeline, channel
decomposition, and robustness tests for all seven elements.
Core orbitals and form factors are obtained from a self-consistent
radial Kohn--Sham solver with Dirac exchange on a uniform grid
($\Delta r = 0.002\;\Bohr$, $r_{\text{max}} = 40\;\Bohr$) for
Na, K, Ca, Mg, Al, and Si.  For Li ($Z = 3$, 1s$^2$ core), the
DFT-LDA solver has a known self-interaction error that makes the
1s orbital $\sim$25\% too diffuse; we instead use the variational
single-exponential form $\psi_{1s} \propto e^{-\alpha r}$ with
$\alpha = Z - 5/16 = 2.6875\;\Bohr^{-1}$, which accounts for
core-core Coulomb screening and gives
$\langle\psi_{\mathrm{var}}|\psi_{\mathrm{HF}}\rangle = 0.9995$
overlap with the numerical Hartree--Fock solution
(see Sec.~\ref{app:lithium} for details).

\subsection{Pipeline}

\textbf{Step~1}: Core excitation energies $\Delta E_c$ are computed
as $\Delta$SCF total energy differences between the $N_c$-electron
ground state and the $(N_c{-}1)$-electron core-hole state.

\textbf{Step~2}: Form factors $f^c_K$
(Sec.~\ref{app:dyson}) are evaluated by radial
quadrature on a uniform grid for each core $s$-orbital.
Core $p$-orbitals are omitted ($f_p \propto j_1(Kr) \to 0$ near
$\Gamma$).  $f^c_K$ is precomputed on a 2000-point grid
($K \in [0, 20]\;\Bohr^{-1}$) and interpolated.

\textbf{Step~3}: For each Bloch state $\ket{\nu\bk}$,
$z^{\mathrm{core}}_{\nu\bk}$ is computed via the closed-form
expression derived in Sec.~\ref{app:dyson} [Eq.~(3) of the main
text] and the QP energy via [Eq.~(2) of the main text].
The full post-SCF correction takes $\sim$0.2~s.

\subsection{Channel decomposition}

Tables~\ref{tab:na_channels} and~\ref{tab:k_channels} give the
per-channel form-factor weights for Na and K\@.
In both cases the outermost $s$-channel dominates; deeper channels
are suppressed by their large $\Delta E_c$.
Table~\ref{tab:trend} summarises $z^{\mathrm{core}}_\Gamma$
across all seven elements, confirming the
$1 - z^{\mathrm{core}} \propto 1/\Delta E_c^{2}$ scaling.

\begin{table}[t]
\caption{\label{tab:na_channels}
Na channel contributions. $f^c_0/\Delta E_c$ is the form-factor
weight entering the coherent sum [Eq.~(3) of the main text].}
\begin{ruledtabular}
\begin{tabular}{lddd}
 & \multicolumn{1}{c}{$\Delta E_c$ (Ha)}
 & \multicolumn{1}{c}{$f^c_0$}
 & \multicolumn{1}{c}{$f^c_0/\Delta E_c$} \\
\hline
1s & 39.91 & -1.05 & -0.026 \\
2s & 2.70  & \multicolumn{1}{c}{$+$1.45} & \multicolumn{1}{c}{$+$0.537} \\
\end{tabular}
\end{ruledtabular}
\end{table}

\begin{table}[t]
\caption{\label{tab:k_channels}
K channel contributions. Same format as Table~\ref{tab:na_channels}.}
\begin{ruledtabular}
\begin{tabular}{ldd}
 & \multicolumn{1}{c}{$\Delta E_c$ (Ha)}
 & \multicolumn{1}{c}{$f^c_0/\Delta E_c$} \\
\hline
1s & \multicolumn{1}{c}{$\sim$200} & \multicolumn{1}{c}{$\sim$0} \\
2s & \multicolumn{1}{c}{$\sim$12}  & \multicolumn{1}{c}{$\sim$0} \\
3s & 1.72     & \multicolumn{1}{c}{$-$0.435} \\
\end{tabular}
\end{ruledtabular}
\end{table}

\begin{table}[t]
\caption{\label{tab:trend}
Core renormalization factor $z^{\mathrm{core}}_\Gamma$ across all
elements studied [Eq.~(3) of the main text].}
\begin{ruledtabular}
\begin{tabular}{lddl}
 & \multicolumn{1}{c}{$z^{\mathrm{core}}_\Gamma$}
 & \multicolumn{1}{c}{$\Delta E_c$ (Ha)}
 & Dom.\ channel \\
\hline
Li & 0.75 & 2.77 & 1s \\
Na & 0.80 & 2.70 & 2s \\
K  & 0.66 & 1.72 & 3s \\
Ca & 0.83 & 2.51 & 3s \\
Mg & 0.91 & 4.07 & 2s \\
Al & 0.96 & 5.69 & 2s \\
Si & 0.98 & 7.57 & 2s \\
\end{tabular}
\end{ruledtabular}
\end{table}

\subsection{Computational details}

All plane-wave DFT calculations use DFTK~\cite{dftk} with LDA
exchange-correlation and Fermi--Dirac smearing at $T = 0.001\;\Ha$.
Table~\ref{tab:comp_details} collects the crystal structures and
pseudopotentials.  All conventional $Z_{\text{val}} = 1$
pseudopotentials (GTH-LDA, HGH-LDA, HGH-PBE) agree on the Li and Na
$\Gamma$-point depths to within $1\;\text{meV}$, confirming that the
static potential is unique for given $Z_{\text{val}}$.

\begin{table}[t]
\caption{\label{tab:comp_details}
Crystal structures and pseudopotentials.}
\begin{ruledtabular}
\begin{tabular}{lccccc}
 & Struct. & $a$ (Bohr) & $Z_v$ & PSP & $E_{\text{cut}}$ (Ha) \\
\hline
Li & BCC & 6.63 & 1 & GTH & 20 \\
Na & BCC & 8.11 & 1 & GTH & 30 \\
K  & BCC & 9.87 & 1 & GTH & 30 \\
Ca & FCC & 10.55 & 2 & HGH & 30 \\
Mg & HCP & 6.07 & 2 & GTH & 30 \\
Al & FCC & 7.65 & 3 & GTH & 30 \\
Si & dia. & 10.26 & 4 & GTH & 30 \\
\end{tabular}
\end{ruledtabular}
\end{table}

\subsection{PBE robustness}
\label{app:pbe_robustness}

To test sensitivity to the static KS calculation, we repeat the
EFT correction using PBE exchange-correlation and PBE
pseudopotentials.  For Na, K, and Ca, PBE provides only semicore
PSPs ($Z_v = 9, 9, 10$); we apply the EFT correction only to the
conduction band.
Table~\ref{tab:pbe_robustness} shows the bandwidth narrowing is
stable: identical to within $0.2$ percentage points for
same-$Z_v$ cases and within $1$--$2$ points otherwise.

\begin{table}[t]
\caption{\label{tab:pbe_robustness}
EFT bandwidth narrowing (\%) with LDA vs PBE.}
\begin{ruledtabular}
\begin{tabular}{lcccc}
 & $Z_v^{\text{LDA}}$ & $Z_v^{\text{PBE}}$
 & LDA (\%) & PBE (\%) \\
\hline
Li & 1 & 3  & 10.1 & 10.4 \\
Na & 1 & 9  & 19.9 & 21.0 \\
K  & 1 & 9  & 34.4 & 36.0 \\
Ca & 2 & 10 & 17.3 & 18.6 \\
\hline
Mg & 2 & 2  & 8.8  & 8.6  \\
Al & 3 & 3  & 4.3  & 4.1  \\
Si & 4 & 4  & 3.7  & 3.7  \\
\end{tabular}
\end{ruledtabular}
\end{table}

\begin{figure*}
\includegraphics[width=\textwidth]{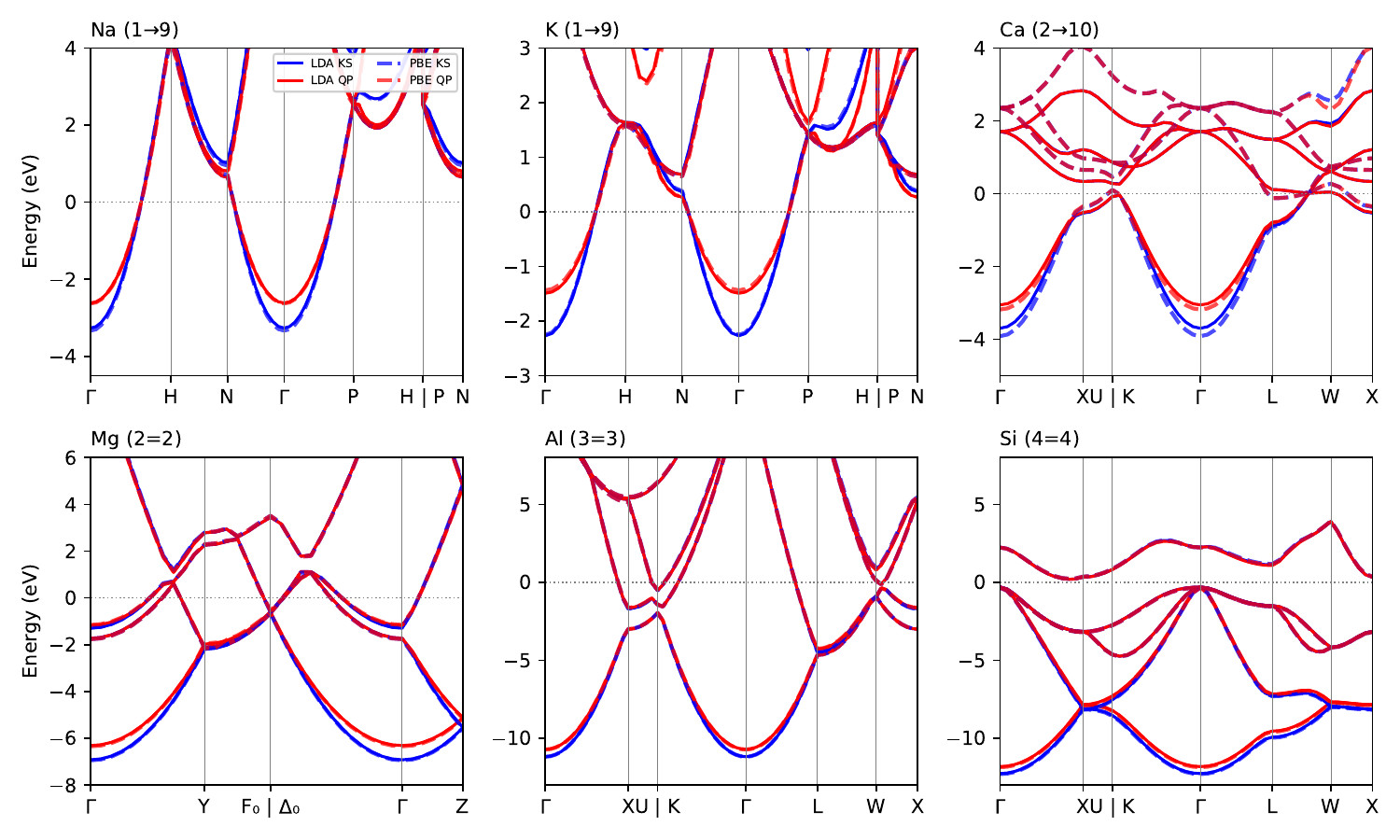}
\caption{\label{fig:pbe_bands}
LDA vs PBE band structures with EFT quasiparticle correction for
seven elements. Solid: LDA; dashed: PBE. Blue: KS; red: QP.}
\end{figure*}

Figure~\ref{fig:pbe_bands} shows the full band structures
overlaying LDA and PBE results with and without the EFT correction.

%=============================================================================
\section{Lithium: explicit EFT potential}
\label{app:lithium}
%=============================================================================

Lithium ($Z = 3$) has a two-electron 1s$^2$ core and one 2s
valence electron, making all EFT integrals analytic.  The
1s orbital is $\psi_{1s}(\br) = (\alpha^3/\pi)^{1/2}\,e^{-\alpha r}$
with $\alpha = Z - 5/16 = 2.6875\;\Bohr^{-1}$ (variational HF for
He-like Li$^+$).  For Li's two-electron core, the DFT-LDA solver
has a known self-interaction error that makes the 1s orbital
$\sim$25\% too diffuse ($\alpha_{\mathrm{eff}} \approx 2.30$ vs.\
$\alpha_{\mathrm{HF}} \approx 2.69\;\Bohr^{-1}$).  The
variational form gives overlap
$\langle\psi_{\mathrm{var}}|\psi_{\mathrm{HF}}\rangle = 0.9995$
with numerical Hartree--Fock.  For heavier atoms (Na--Si) the
multi-electron core is well described by DFT-LDA and this issue
does not arise.

The relevant atomic quantities are:
orbital energy $\epsilon_{1s} = \alpha^2/2 - Z\alpha = -4.449\;\Ha$,
self-Coulomb integral $J = 5\alpha/8 = 1.680\;\Ha$,
Li$^+$ ground-state energy
$E_0 = 2\epsilon_{1s} + J = -7.219\;\Ha$,
and core excitation energy
$\Delta E_{1s} = E_1 - E_0 = 2.770\;\Ha$.
The Coulomb potential from one 1s electron is
\begin{equation}\label{eq:u}
    u(r) \equiv V_{H,1s}(r) = \frac{1 - e^{-2\alpha r}}{r}
    - \alpha\, e^{-2\alpha r},
\end{equation}
with $u(0) = \alpha$ and $u(r) \to 1/r$ at large $r$.

\subsection{Four-term formula and form factor}

Substituting the 1s orbital into Eq.~\eqref{eq:dvpp_ion}:
\begin{equation}\label{eq:dvpp_li}
\begin{aligned}
    &\delta V_{pp}^{(\text{ion})}(\br,\br') \\
    &= 2u(r)\,\delta(\br - \br')
       - \frac{\alpha^3}{\pi}\frac{e^{-\alpha(r+r')}}{|\br - \br'|}  \\
    &\quad - \frac{\alpha^3}{\pi}e^{-\alpha(r+r')}\bigl[u(r)+u(r')-J\bigr] \\
    &\quad + \frac{\alpha^3}{\pi}e^{-\alpha(r+r')}
             \frac{\bigl(u(r)-J\bigr)\bigl(u(r')-J\bigr)}{i\omega + \Delta E_{1s}}.
\end{aligned}
\end{equation}
The analytic form factor [Eq.~(7)
of the main text] evaluates to
\begin{equation}\label{eq:AK}
    f^{1s}_K = 8\sqrt{\pi/\alpha}\, A_K, \quad
    A_K = \frac{Q^2 + 33}{(Q^2+1)(Q^2+9)^2}
      - \frac{J}{\alpha}\frac{1}{(Q^2+1)^2},
\end{equation}
where $Q = K/\alpha$.  For Li's single 1s channel,
$z^{\mathrm{core}}_{\nu\bk}$ reduces to
$1/(1 + |F^{1s}_{\nu\bk}|^2/\Delta E_{1s}^2)$, giving
$z^{\mathrm{core}}_\Gamma \approx 0.75$.

\subsection{Nonlocal pseudopotential}

The 1s exchange kernel has different strengths in each $l$ channel:
for $l = 0$ both exchange and projection contribute; for $l \geq 1$
only exchange survives (projection vanishes by angular selection).
We solve the radial Schr\"odinger equation on a uniform grid
($dr = 0.01\;\Bohr$, $r_{\mathrm{max}} = 30\;\Bohr$),
apply Troullier--Martins pseudization ($r_c = 3.0\;\Bohr$),
and construct a Kleinman--Bylander PSP with
$V_{\mathrm{local}} = V^{l=1}$ and a single $l = 0$ projector.

\subsection{Band structure comparison}

\begin{figure}[t]
\centering
\includegraphics[width=\columnwidth]{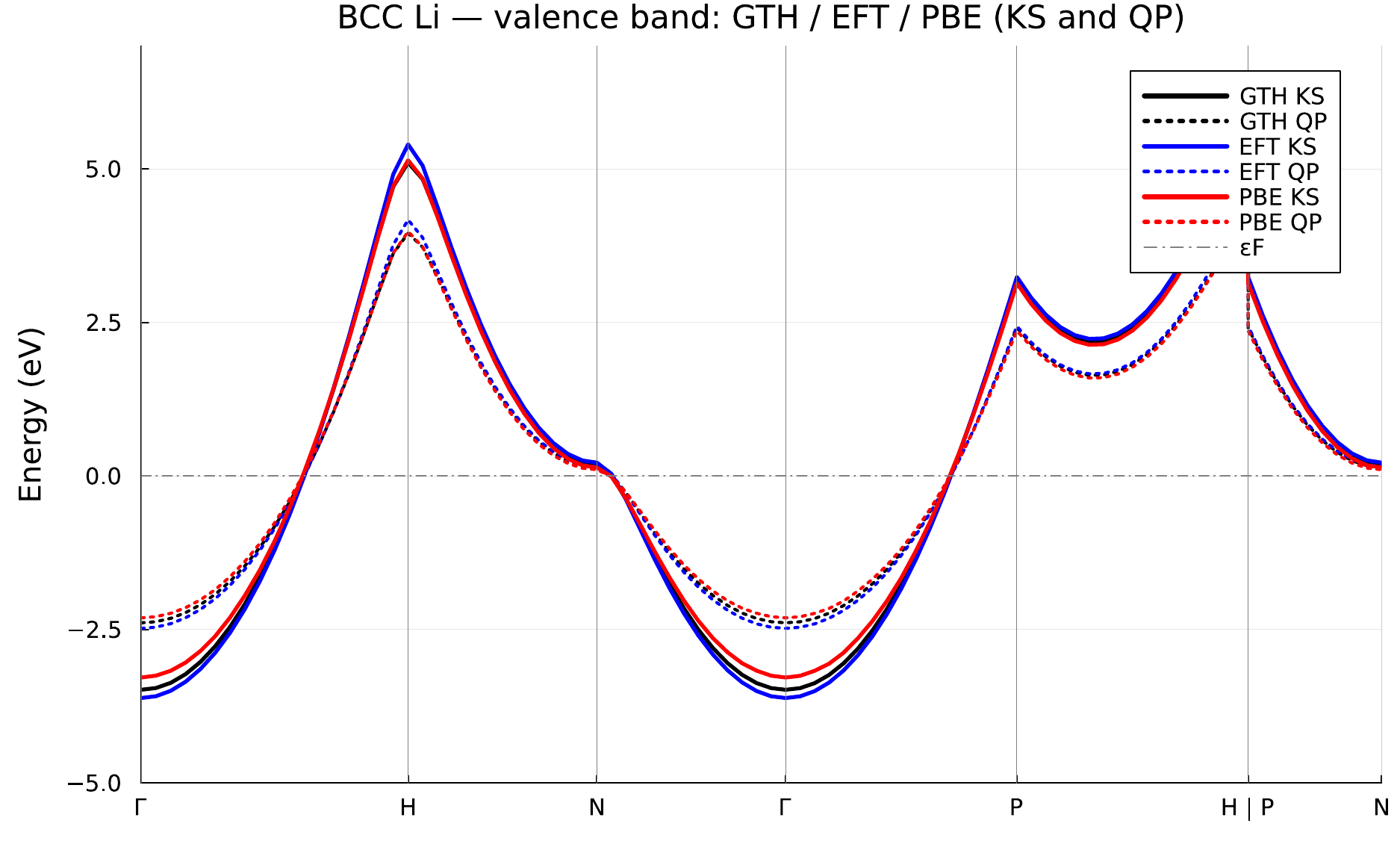}
\caption{\label{fig:li_bands}
Li BCC band structure along N--$\Gamma$--H.  Blue: KS; red: EFT QP.
Solid: GTH-LDA ($Z_v=1$); dashed: EFT nonlocal KB PSP;
dotted: PBE ($Z_v=3$).}
\end{figure}

Figure~\ref{fig:li_bands} compares the Li BCC valence band from
three static PSPs---conventional GTH-LDA ($Z_v = 1$), EFT nonlocal
KB, and PBE ($Z_v = 3$)---each with and without the QP correction.

The EFT PSP gives a $\Gamma$-point depth of $-3.81\;\eV$ vs.\
GTH $-3.48\;\eV$ ($\sim$10\% difference), confirming that the
static EFT potential recovers conventional PSP physics.
Using variational HF ($\alpha = 2.6875$) instead of hydrogenic
($\alpha = 3$) reduces the $\Gamma$-point error from $0.33$ to
$0.13\;\eV$.  The QP correction compresses the bandwidth by
$\sim$25\% regardless of the static PSP, confirming that
$z^{\mathrm{core}}$ is a post-SCF property insensitive to the
choice of static potential.

\end{document}